\documentclass[11pt,twocolumn]{IEEEtran}
\usepackage{cite}

\title{Spatial sampling and beamforming for spherical microphone arrays}
\author{Boaz Rafaely \\ Department of Electrical and Computer Engineering \\ Ben-Gurion University of the Negev. Israel, br@bgu.ac.il, 13 April 2008
\thanks{This research was supported in part by
THE ISRAEL SCIENCE FOUNDATION (grant No. 155/06).} }

\begin{document}
%\ninept
%
\maketitle
\begin{abstract}
Spherical microphone arrays have been recently studied for spatial
sound recording, speech communication, and sound field analysis
for room acoustics and noise control. Complementary theoretical
studies presented progress in spatial sampling and beamforming
methods. This paper reviews recent results in spatial sampling
that facilitate a wide range of spherical array configurations,
from a single rigid sphere to free positioning of microphones. The
paper then presents an overview of beamforming methods recently
presented for spherical arrays, from the widely used delay-and-sum
and Dolph-Chebyshev, to the more advanced optimal methods,
typically performed in the spherical harmonics domain.
\end{abstract}
%
%\begin{keywords}
%Microphone array, spherical harmonics, spatial sampling,
%beamforming, array robustness.
%\end{keywords}
%

%%%%%%%%%%%%%%%%%%%%%%%%%%%%%%%%%%%%%%%%%%%%%%%%%%%%%%
\section{Introduction}
\label{sec:intro}
%%%%%%%%%%%%%%%%%%%%%%%%%%%%%%%%%%%%%%%%%%%%%%%%%%%%%%

The growing interest in spherical microphone arrays can probably
be attributed to the ability of such arrays to measure and analyze
three-dimensional sound fields in an effective manner. The other
strong point of these arrays is the ease of array processing
performed in the spherical harmonics domain. The recent papers by
Meyer and Elko \cite{meyer2002} and Abhayapala and Ward
\cite{abhayapala2002}, presented the use of spherical harmonics in
array processing and offered some basic array configurations. The
studies that followed, some of which are presented in this paper,
showed theoretical developments, experimental investigations, and
signal processing methods.

This paper presents an overview of some recent results in two
aspects of spherical microphone arrays, namely spatial sampling
and beamforming. The former facilitates improved array
configurations, while in the latter a wide range of beamforming
methods available in the standard array literature are adapted for
spherical arrays.

\section{Spherical array processing}\label{sec:proc}

The theory of spherical array processing is briefly outlined in
this section. Consider a sound field with pressure denoted by
$p(k,r,\Omega)$, where $k$ is the wave number, and
($r,\Omega)\equiv(r,\theta,\phi)$ is the spatial location in
spherical coordinates \cite{williams1999}. The spherical Fourier
transform of the pressure is given by \cite{williams1999}:
\begin{equation} \label{SFT}
p_{nm}(k,r) = \int_{\Omega\in S^2} p(k,r,\Omega) Y_n^{m^*}(\Omega)
\, d\Omega,
\end{equation}
with the inverse transform relation:
\begin{equation} \label{ISFT}
p(k,r,\Omega) = \sum_{n=0}^\infty \sum_{m=-n}^{n} p_{nm}(k,r)
Y_{n}^{m}(\Omega),
\end{equation}
where $Y_{n}^{m}(\Omega)$ is the spherical harmonics
\cite{williams1999} of order $n$ and degree $m$. The sound field
can be represented by potentially infinite number of plane waves
having spatial amplitude density $a(k,\Omega_0)$, with $\Omega_0$
denoting the arrival directions, in which case $p_{nm}$ can be
written as \cite{rafaely2004}:
\begin{equation} \label{pnm1}
p_{nm}(k,r) = a_{nm}(k) b_n(kr),
\end{equation}
where $b_n(kr)$ depends on the sphere boundary and has been
presented for rigid sphere, open sphere, and other array
configurations \cite{balmages2007}, and $a_{nm}(k)$ is the
spherical Fourier transform of $a(k,\Omega_0)$.

Spherical array processing typically includes a first stage of
approximating $p_{nm}$, or $a_{nm}$, followed by a second
beamfroming stage performed in the spherical harmonics domain
\cite{meyer2002,abhayapala2002,rafaely2005SAP}. For the first
stage, we can write:
\begin{equation} \label{pnm2}
\hat{p}_{nm}(k) = \sum_{j=1}^M c_{nm}^j(k) p(k,r_j,\Omega_j).
\end{equation}

Array input is measured by $M$ pressure microphones located at
$(r_j,\Omega_j)$, and is denoted by $p(k,r_j,\Omega_j)$. Array
order is $N$, typically satisfying $(N+1)^2\le M$
\cite{rafaely2005SAP}. Coefficients $c_{nm}^j$, which may be
frequency-dependant, are selected to ensure accurate approximation
of the integral in (\ref{SFT}) by the summation in (\ref{pnm2}).
Substituting (\ref{ISFT}) and (\ref{pnm1}) in (\ref{pnm2}) we can
write:
\begin{equation} \label{anm}
\hat{a}_{nm}(k) = \sum_{n'=0}^\infty \sum_{m'=-n'}^{n'} a_{n'm'}
\sum_{j=1}^M c_{nm}^j(k) b_{n'}(kr_j) Y_{n'}^{m'}(\Omega_j).
\end{equation}

Ideally, we would like the following to hold:
\begin{equation} \label{orth}
\sum_{j=1}^M c_{nm}^j(k) b_{n'}(kr_j) Y_{n'}^{m'}(\Omega_j) =
\delta_{nn'}\delta_{mm'},
\end{equation}
where $\delta_{nn'}$ is the Kronecker delta function, in which
case we get $\hat{a}_{nm}=a_{nm}$. Typically, this will hold for
$n',n\le N$, and so the approximation will be accurate for
order-limited functions. The orthogonality condition in
(\ref{orth}) can be written in a matrix form as
\cite{rafaely2008SAP}:
\begin{equation} \label{orthvec}
\mathbf{B} \mathbf{c}_{nm} = \mathbf{I}_{nm},
\end{equation}
where $\mathbf{c}_{nm}$ is the $M \times 1$ vector of the
coefficients $c_{nm}^j$, $\mathbf{I}_{nm}$ is an $(N+1)^2 \times
1$ zero vector with a single element of one, and $\mathbf{B}$ is
the $(N+1)^2 \times M$ matrix with elements $b_n(kr_j)
Y_n^m(\Omega_j)$. A good choice of microphone locations should
produce a mall condition number of matrix $\mathbf{B}$, therefore
providing good array robustness in terms of white-noise-gain
\cite{rafaely2008SAP}. Once $p_{nm}$ or $a_{nm}$ have been
measured, array output $y(k)$ can be computed by
\cite{rafaely2005SAP,rafaely2008SPL}:
\begin{equation} \label{y1}
y(k) =  \sum_{n=0}^N \sum_{m=-n}^n d_{nm}\, \hat{a}_{nm}(k),
\end{equation}
where $d_{nm}$ controls the beam pattern and $N$ is the array
order. In the more common case of a beam pattern that is symmetric
around the look direction, $d_{nm}=d_n Y_n^m(\Omega_l)$
\cite{meyer2002}, where $\Omega_l$ is the array look direction.
However, in this paper we will also consider the more general beam
pattern design and steering \cite{rafaely2008SPL}.

\section{Spatial Sampling}

Arrays with high robustness require maintaining (\ref{orth}) while
ensuring a low condition number for matrix $\mathbf{B}$. In the
particular case that the microphones are arranged on the surface
of a single sphere, this requirement is decoupled into two
conditions. The first requires maintaining (\ref{orth}), but
without the term $b_n$ in the equation, as now $r$ is constant and
no longer a function of $j$. There are various sampling
configurations that satisfy this requirement \cite{rafaely2005SAP,
li2007}. The second condition simply states that $b_n(kr)$ should
avoid small values, since then an entire row in matrix
$\mathbf{B}$ will be near zero, or $p_{nm}$ may not be measurable,
as suggested by (\ref{pnm1}). Several array configurations and
their robustness are presented in this section.

\subsection{Single Open Sphere}
Microphones arranged around open spheres have been employed
previously \cite{gover2002}. This configuration is useful in
arrays of large radius where a rigid sphere is impractical, or
when a scanning microphone array is employed
\cite{rafaely2007jasa}. In this case we get:
\begin{equation}
b_n(kr) = 4 \pi i^n j_n(kr)
\end{equation}
where $j_n(kr)$ is the spherical Bessel function, satisfying
$b_n(kr)=0$ at some values of $kr$ and $n$. At these values matrix
$\mathbf{B}$ has a line of zeros and its inversion becomes
unstable. Several solutions to this problem have been reported and
are discussed in the subsections that follow.

\subsection{Single rigid Sphere}
A single rigid sphere \cite{meyer2002} is the preferred array
configuration due to its simplicity, ease of microphone mounting,
and numerical robustness, with:
\begin{equation}
b_n(kr) = 4 \pi i^n \left(j_n(kr)-\frac{j'_n(kr_0)}{h'_n(kr_0)}
h_n(kr)\right).
\end{equation}
In this case $b_n(kr)$ avoids low values for $kr>n$, ensuring a
wide operating frequency range.

\subsection{Hemisphere Configuration}
Li and Duraiswami presented an interesting configuration
\cite{li2005}, where only a hemispherical surface of microphones
is required when the array is mounted against a rigid surface,
taking advantage of the symmetry of the sound field. This
configuration offers a reduced number of microphones and may be
attractive in a wide range of applications.

\subsection{Open Sphere with Cardioid Microphones}
An open sphere with cardioid microphones, which combine pressure
and pressure gradient measurements, produces \cite{balmages2007}:
\begin{equation}
b_n(kr) = 4\pi i^n [j_n(kr) - i j'_n(kr)],
\end{equation}
which also provides a solution to the problem of the zeros in the
spherical Bessel function. However, cardioid microphones are less
common and may induce uncertainty due to imperfect directivity and
mounting.

\subsection{Dual Sphere Configuration}
A dual-sphere configuration has been introduced to overcome the
numerical conditioning of open-sphere arrays, but by using
pressure microphones which are more common. In this case:
\begin{equation}
b_n(kr) = 4\pi i^n [(1-\beta_n(kr)) j_n(kr) + \beta_n(kr)
j_n(k\alpha r)],
\end{equation}
where $\beta_n(kr)$ is a sphere-selecting parameter, and the two
radii are $r$ and $\alpha r$. A derivation for the optimal value
of $\alpha$ for given array order and frequency range has also
been recently presented \cite{balmages2007}. The method has been
experimentally applied to room acoustics studies
\cite{rafaely2007jasa}.

\subsection{Spherical Shell Configuration}
The dual-sphere configuration requires twice as many microphones
compared to a single-sphere configuration, and so in a recent
paper \cite{rafaely2008SAP}, a configuration in which the
microphones are placed within the volume of a spherical shell has
been proposed. This configuration was shown to overcome the
numerical conditioning problem but without using additional
microphones. It has been shown that in this case a low condition
number for matrix $\mathbf{B}$ can be achieved.

\subsection{Free Sampling Configuration}
The framework presented in \cite{rafaely2008SAP} facilitates the
design of free sampling configurations, by placing microphones
anywhere in three-dimensional space, and optimizing or fine-tuning
their position to achieve minimal condition number of matrix
$\mathbf{B}$. This framework allows the study of new array
configurations. The method can be viewed as an extension to
another recent study where flexibility in microphone placement was
introduced, but limited to the surface of a single rigid sphere
\cite{li2007} .

\section{Beamforming}

Once the sampling configuration has been selected, array
processing, or beamforming, can be performed. Beamformer design is
the process of selecting array weights $d_n$ or $d_{nm}$ to
achieve some array performance objectives. This section reviews
some recent results in spherical array beamforming.

\subsection{Regular Beam Pattern}
A beam pattern which is symmetric around the look direction, i.e.
depends only on the angle away from the look direction, can be
designed by selecting $d_{nm}=d_n Y_n^m(\Omega_l)$ in (\ref{y1}).
In the simple case that $d_n=1$, the so-called regular beam
pattern \cite{li2007}, or plane-wave decomposition array
\cite{rafaely2004}, is achieved. In this case $\hat{a}_{nm}
\rightarrow a_{nm}$ as $N \rightarrow \infty$, resulting in
perfect decomposition of the sound field into plane-wave
components.

\subsection{Delay and Sum Beam Pattern}
A common beamforming patten is achieved by a delay-and-sum array,
which exhibits the attractive property of a constant
white-noise-gain. Such an array has been developed for spherical
microphone arrays and was shown to satisfy $d_n=|b_n(kr)|^2$
\cite{rafaely2005SPL}. Note that a delay-and-sum array, typically
implemented for sensors in free-field by applying weights that
correspond to the differences in delay between the sensors
relative to the plane wave arrival direction, is applicable here
for both open sphere and rigid sphere array configurations!

\subsection{Dolph-Chebyshev Beam Pattern}
Another beam pattern which is useful in practice is the
Dolph-Chebyshev beam pattern \cite{vantrees2002}. This beam
pattern achieves the lowest side-lobe level for a given array
order and main-lobe width, or alternatively the narrowest main
lobe for a given array order and side-lobe level. In a recent
study \cite{koretz2008}, a formulation for directly calculating
values of $d_n$ to produce Dolph-Chebyshev beam patterns for
spherical arrays has been presented. Application of the method for
room acoustics analysis has also been recently presented
\cite{rafaely2007ISRA}.

\subsection{Optimal Beam Pattern}
Li and Duraswami presented an optimal spherical microphone array
design framework, where design objectives such as directivity and
robustness can be considered \cite{li2007}. The weights $d_n$ are
therefore the result of an optimization procedure, and are optimal
under the given conditions. This method can provide arrays that
are useful in practice due to the ability to constrain the
white-noise-gain, therefore producing arrays that are robust to
uncertainties and noise.

\subsection{Beam Pattern with Desired Multiple Nulls}
In some cases, such as plane wave decomposition for room
acoustics, the array is required to enhance a plane wave from a
given direction while suppressing plane waves from other
directions. One way to achieve such suppression is by placing
nulls in the beam pattern pointing to undesired wave directions
\cite{rafaely2008ICASSP}. In this case the resulting beam pattern
is not necessarily symmetric around the look direction, such that
array weights are represented by $d_{nm}$ and not $d_n$.

\subsection{Non-Symmetric Beam Pattern and its Steering}
Beam patterns that are not symmetric around the look direction may
also be useful in other applications. A challenge that arises with
such beam patterns is beam steering. Steering of beam patterns
that are symmetric around the look direction is performed simply
by changing the value of $\Omega_l$ in $d_{nm}=d_n
Y_m^m(\Omega_l)$, keeping the beam shape, controlled by $d_n$,
unchanged. This attractive property of decoupling of beam pattern
from its steering was recently presented for more general beam
patterns \cite{rafaely2008ICASSP}. The steering is achieved by a
weighted summation of $d_{nm}$ with
$D_{m'm}^n(\alpha,\beta,\gamma)$, which is the Wigner-D function
holding the steering, or beam rotation angles.

\subsection{Near-field Beam Pattern}
The beam patterns described above are all designed under the
assumption that the sources are in the far field. In some
applications, such as hand-free speech communication and music
recording, the sources may be close to the array, and so the plane
wave assumption may no longer hold. Meyer and Elko
\cite{meyer2006} presented a close-talk array for near field
sources, where the distance of the source is approximated from
array data. Recently, Fisher and Rafaely \cite{fisher2008ICASSP}
presented a study of the near field spherical array, investigating
the extension of the near field zone, and controlling the beam
pattern along the radius.

\subsection{Direction of arrival estimation}
The task of estimating the direction of arrival of a source has
been of interest for a wide range of applications
\cite{vantrees2002}. Teutsch has recently presented a subspace
method for direction-of-arrival estimation for spherical
microphone arrays, performed in the spherical harmonics domain
\cite{teutsch2007}. The method extends similar algorithms
developed for linear and circular arrays.

\section{Conclusion}
This paper presented an overview of recent studies of spherical
microphone arrays, concentrating on two topics, namely microphone
positioning, or spatial sampling, and beamforming. The results
presented in this paper can be used to design spherical microphone
arrays and implement array processing in a wide range of
applications.

%\section{REFERENCES}
%\label{sec:ref}

%List and number all bibliographical references at the end of the paper.  The references can be numbered in alphabetic order or in order of appearance in the document.  When referring to them in the text, type the corresponding reference number in square brackets as shown at the end of this sentence \cite{C2}.


\bibliographystyle{IEEEtran}
\bibliography{HSCMA2008Rafaely}

%\begin{figure}
%\centering
%\includegraphics[width=8.5cm, angle=0]{fig_regular.eps}
%\caption{Regular beampattern.} \label{fig_regular}
%\end{figure}

\end{document}